\documentclass[english,a4paper]{article}
\usepackage[T1]{fontenc}
\usepackage[cp1250]{inputenc}
\usepackage{amssymb}
\usepackage{hyperref}
\usepackage{amsfonts}
\usepackage{graphicx}
\usepackage{revsymb}
\usepackage{dcolumn}
\usepackage{bm}

\def\phi{\varphi}
\def\epsilon{\varepsilon}

\newcommand{\Exp}[1]{{\rm e}^{#1}}

\newcommand{\bec}{\begin{center}}
\newcommand{\enc}{\end{center}}
\newcommand{\be}{\begin{equation}}
\newcommand{\ee}{\end{equation}}
\newcommand{\bmi}{\begin{minipage}}
\newcommand{\emi}{\end{minipage}}

\newcommand{\bi}{\begin{itemize}}
\newcommand{\ei}{\end{itemize}}
\newcommand{\ba}{\begin{array}}
\newcommand{\ea}{\end{array}}

\newcommand{\lgr}{\left\{}

\newcommand{\rd}{\right.}

\everymath{\displaystyle}

\begin{document}
%\title{Optimal signal to noise ratio per unit time of a spin 1/2 particle: The Ernst angle solution}
\title{Optimal control of the signal to noise ratio per unit time for a spin 1/2 particle}
\author{Marc Lapert\footnote{Department Chemie, Technische Universit{\"a}t
M{\"u}nchen, Lichtenbergstrasse 4, 85747 Garching, Germany}, E. Ass\'emat, S. J. Glaser\footnote{Department Chemie, Technische Universit{\"a}t
M{\"u}nchen, Lichtenbergstrasse 4, 85747 Garching, Germany} and D. Sugny\footnote{Laboratoire Interdisciplinaire
Carnot de Bourgogne (ICB), UMR 5209 CNRS-Universit\'e de
Bourgogne, 9 Av. A. Savary, BP 47 870, F-21078 DIJON Cedex,
FRANCE, dominique.sugny@u-bourgogne.fr}}

\maketitle

\begin{abstract}
We investigate the maximum signal to noise ratio per unit time that can be achieved for a spin 1/2 particle subjected to a periodic pulse sequence. Optimal control techniques are applied to design the control field and the position of the steady state, leading to the best signal to noise performance. A complete geometric description of the optimal control problem is given in the unbounded case. We show the optimality of the well-known Ernst angle solution, which is widely used in spectroscopic and medical imaging applications, over a large control space allowing use of shaped pulses.
\end{abstract}
%\pacs{32.80.Qk,37.10.Vz,78.20.Bh}
%\maketitle

\section{Introduction}
Optimal control theory has been developed for engineering applications \cite{pont,bonnardbook}, but it is nowadays a key tool to quantitatively analyze the dynamics of complex systems including such where the quantum effects are predominant \cite{rice,reviewQC}. According to the applications under concern, optimal control techniques allow us by means of shaped pulses to maximize the yield of a given observable \cite{reviewQC} or to determine the minimum control time to reach the target state \cite{lapertglaser,thomas,boscain,assemat,damping,optimaging,khaneja}. In this paper, we consider a non standard control problem for which the quantum system evolves periodically under repeated application of the pulse sequence. This periodic controlled dynamics is crucial in many different domains where the signal to noise ratio (SNR) increases with the number of scans. In this case, the initial and final state of the dynamics, called a steady state, is not known and depends on the used control field.

We show how to optimize the SNR per unit time in a specific quantum system, namely a spin 1/2 particle in Nuclear Magnetic Resonance (NMR) \cite{spin}, whose dynamics can be controlled through a radio-frequency magnetic field \cite{carr,haacke,scheffler}. This problem finds direct applications in Magnetic Resonance Imaging (MRI), where the SNR is one of the crucial features of fast imaging techniques \cite{bookimaging}. In particular, this analysis shows  the optimality of the Ernst angle solution \cite{ernstangle}, which is well-established and widely used in spectroscopic and medical imaging applications, in the general case of unbounded control including finite-amplitude  shaped pulses. Note that the corresponding control law was established by considering only delta-pulses, a constraint which is relaxed in this work. In addition to its interest in spin dynamics, our approach paves the way to a systematic use of optimal control techniques in other domains beyond NMR or Electron Paramagnetic Resonance (EPR) \cite{EPR} where it is also desirable to maximize the SNR for a given measurement time.

The paper is organized as follows. The model system is presented in Sec. \ref{sec2} with an explicit derivation of the figure of merit used to define the optimal control problem. Section \ref{sec3} focuses on the different control sequences applied according to the position of the steady state. Section \ref{sec4} is devoted to the computation of the optimal steady state. Conclusion and prospective views are given in a final section \ref{sec5}. Technical computations are reported in the Appendices \ref{appa} and \ref{appb}.

%More specifically, we will consider the optimization of the SNR of a spin 1/2 particle in Nuclear Magnetic Resonance (NMR) \cite{spin}, . After a transient relaxation process which will not be described in this paper, the dynamical system reaches a steady state. . Using tools of geometric optimal control theory \cite{bonnardbook,boscainbook}, we show how to determine the steady state position and the corresponding optimal pulse sequence in order to maximize the SNR.

%The paper is organized as follows. After having introduced the model system, we give a complete geometric description of the steady state problem in the unbounded case. As a byproduct of this computation, we show the optimality of the Ernst angle solution established in the sixties. We discuss in the conclusion some possible extensions of this preliminary work.

\section{The model system}\label{sec2}
\begin{figure}[htbp]
\centering\includegraphics[scale=0.6]{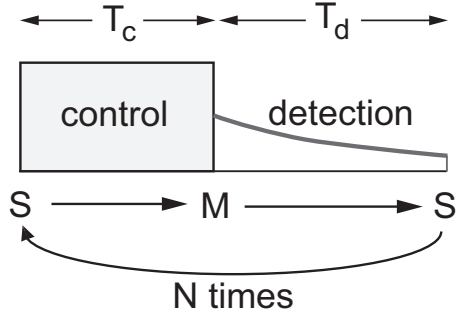}
\caption{\label{fig0} Schematic description of the cyclic controlled process (see the text for the definition of the different points and parameters).}
\end{figure}

Many different aspects and control scenarios could be considered for the optimization of the SNR of a spin 1/2 particle. In this paper, we will restrict the study to a relatively simple but realistic scenario. The process inspired from MRI problems can be schematically described as follows. A given experimental block made of a pulse sequence and of a detection period with free relaxation dynamics is repeated many times. This control process is schematically illustrated in Fig. \ref{fig0}. After a transient relaxation process, we assume that the system has reached the steady state condition, either by use of shaped pulse designed by optimal control or by a repeated application of the experimental cycle. In this periodic regime, the accumulated signal strength increases linearly as a function of the block number, while the noise and therefore the SNR vary as the square root. A multitude of cycles are then practically applied until a satisfactory level of SNR is achieved. The quality of the process can be estimated with the aid of a figure of merit describing the SNR per unit time (see below for an explicit derivation). From an optimal control perspective, we are therefore interested in finding the pulse sequence which maximizes this figure of merit.

To be more specific, we consider a homogeneous ensemble of uncoupled spin 1/2 particles. In a given rotating frame, the equation of motion for the ensemble irradiated on resonance is given by \cite{spin}:
\begin{eqnarray*}
&& \dot{M}_x=-2\pi M_x/T_2+\omega_y M_z \\
&& \dot{M}_y=-2\pi M_y/T_2-\omega_x M_z  \\
&& \dot{M}_z= 2\pi(M_0-M_z)/T_1+\omega_x M_y-\omega_y M_x, \\
\end{eqnarray*}
where $\vec{M}=(M_x,M_y,M_z)$ is the Bloch vector, $(\omega_x,\omega_y)$ the two components of the radio-frequency magnetic field along the  $x$- and the $y$- directions and $T_1$ and $T_2$ the longitudinal and transverse relaxation parameters. $M_0$ is the magnitude of the Bloch vector at thermal equilibrium. Using the symmetry of revolution of the system around the $z$- axis, we can restrict the problem to only one control field \cite{BS,BCS}. In the following of the paper, we will assume without loss of generality that $\omega_y=0$, which implies that the $x$- coordinate of the Bloch vector is not coupled to the other components. Introducing the normalized coordinates $y=M_y/M_0$ and $z=M_z/M_0$, the dynamics which takes place in the $(y,z)$- plane is ruled by the following set of equations:
\begin{equation}\label{eq1}
\dot{y}=-2\pi y/T_2-uz;~\dot{z}=2\pi(1-z)/T_1+uy,
\end{equation}
where the control field is $u=\omega_x$. We assume that the elementary cycle of the experimental process is repeated $N$ times, with $N\gg 1$. We denote by $T$, the total time of the process, by $T_d$ the duration of one measurement and by $T_c$ the time of the pulse sequence. The detection period $T_d$ and the total time $T$ are fixed by the experimental setup. The different parameters satisfy the relation $T=N(T_d+T_c)$. As can be seen in Fig. \ref{fig0}, the steady state $S$ is defined as the state at the end of the detection (i.e. at the beginning of the control period $T_c$) and $M$ as the point where the measurement process starts. The coordinates of the steady state and of the $M$ point are denoted by $(y_s,z_s)$ and $(y_m,z_m)$, respectively. The polar coordinates $(r_s,\theta_s)$ and $(r_m,\theta_m)$ will be also used in the following. The measured signal in NMR is the transverse component of the Bloch vector $\vec{M}$, i.e. the $y_m$- component \cite{spin}. The accumulated signal grows linearly with the number of repetitions of the elementary block, while the detected noise (we assume a white noise) increases with the square root. The SNR, denoted by $R$, is then proportional to:
$$
R=\frac{Ny_m}{\sqrt{N}}=\frac{\sqrt{T}}{\sqrt{T_d+T_c}}y_m,
$$
which leads to the definition of the quality factor $Q=\frac{y_m}{\sqrt{T_d+T_c}}$. Since $T$ is fixed, maximizing the SNR $R$ is equivalent to maximizing the SNR per unit time $Q$. Normalizing the time $t$ by $T_d$ and setting $\Gamma=2\pi T_d/T_2$ and $\gamma=2\pi T_d/T_1$, the dynamical system becomes :
\begin{equation}\label{eq1new}
\dot{y}=-\Gamma y -u z;~\dot{z}=\gamma(1-z)+u y,
\end{equation}
and the figure of merit is given by $Q(y_m,z_m)  = y_m/\sqrt{1+T_{c}}$, $T_c$ being a function of $(y_m,z_m)$. The maximum possible value of $Q$ is 1 for $y_m=1$ and $T_c=0$, but this upper bound cannot be reached by any solution of the problem.

The control process is aimed at maximizing the figure of merit $Q$ by designing the optimal control sequence, but also the position of the steady state, since this latter depends on the used control sequence. We are therefore faced with a non trivial control problem in which the initial and the final states are not fixed. We use the following brute-force approach to answer geometrically this question in the unbounded case, for which there is no constraint on the amplitude of the control field. Let us fix a point $S$ of the $(y,z)$- plane. The $M$ point can be straightforwardly determined by integrating backwards in time Eq.~(\ref{eq1new}) with $u=0$. We then search for the time-minimum control field, which drives the system from $S$ to $M$. The general solution of the time-optimal control problem has been given in a series of papers, both for the bounded \cite{lapertglaser,BS,BCS} and the unbounded cases \cite{magic}. It can be shown that the structure of the time-optimal control law can be mainly described by two geometric objects of the Bloch ball which play a central role in the present analysis: the magic plane of equation $z=z_0=\frac{-\gamma}{2(\Gamma-\gamma)}$ and the $z$-axis. A brief description of these structures is given in the Appendix \ref{appa}.
\section{Classification of the steady state control}\label{sec3}
In order to construct the $Q$ surface, we need to classify the structure of the time-optimal control field associated with each point of the $(y,z)$- plane. By analogy with a standard time-optimal synthesis \cite{boscainbook}, such a classification is called in this paper a steady state synthesis. Since we consider unbounded control, the rotation from $\theta_s$ to $\theta_m$ can be done instantaneously trough the use of $\delta$-pulse, thus the problem reduces to a radius transfer problem from $r_s$ to $r_m$. Based on this remark and knowing that the magic plane and the $z$-axis are the sets where shrinking and growing of radius are maximum \cite{magic}, five different pulse sequences can be identified for controlling in minimum time the system. A complete derivation of the control laws can be made as follows. Different examples of the corresponding trajectories are displayed in Fig. \ref{fig5} (see also the movies in the supplementary material).

During the detection period, the radius grows ($r_m<r_s$) for some points and shrinks ($r_m>r_s$) for others, where $r_m=\sqrt{y_m^2+z_m^2}$ and $r_s=\sqrt{y_s^2+z_s^2}$ are the radial coordinates of the $S$ and $M$ points. In the second situation, we know from the previous paragraph that the fastest way to increase the radius from $r_s$ to $r_m$ is to reach the set $(y=0,z>0)$. For this purpose, a first bang (i.e. here a $\delta$- pulse of a given angle) should be used to move the state of the system along an arc of circle from S to the point of coordinates $(y=0,z=r_s)$. After a free relaxation along the $z$- axis from $z=r_s$ to $z=r_m$, a second bang is then used to reach the $M$ point (blue trajectory in Fig. \ref{fig5}). In the case $r_m<r_s$, different control laws can occur. If $\gamma/\Gamma>2/3$ then the magic plane does not intersect the Bloch ball. The time optimal solution is made of a bang pulse to reach the set where the radial speed is maximum, i.e. the set $(y=0,z<0)$, to decrease the $z$- component from $z=-r_s$ to $z=-r_m$ and then a second bang pulse is used to recover the $M$ point (red trajectory in Fig. \ref{fig5}). If $\gamma/\Gamma<2/3$, the fastest way to reduce the radius is to follow the magic plane. Three sub-cases can be encountered. If the radius $r_s$ is larger than $|z_0|$ and $r_m$ is lower than $|z_0|$ then the control law is the concatenation of a bang pulse to reach the magic plane, followed successively by a trajectory along the magic plane up to the point $(y=0,z=z_0)$, by a zero control along the $z$- axis to reach the point of coordinate $z=-r_m$, and another bang pulse to bring the spin to $M$ (orange trajectory in Fig. \ref{fig5}). If $r_s$ and $r_m$ are larger than $z_0$, we should go to the magic plane, shrink the radius up to $r_m$ and use a second bang to attain $M$ (green trajectory in Fig. \ref{fig5}). If $|z_0|$ is bigger than $r_m$ and $r_s$, the magic plane cannot be used and the line $(y=0,z_0<z<0)$ has to be used to shrink the radius (red trajectory in Fig. \ref{fig5}).

This analysis leads to the five possible control structures which can be summarized as follows:
\bi
\item $r_s < r_m$ $\rightarrow$ $BS_{v>0}B$
\item $r_s = r_m$ $\rightarrow$ $B$
\item $r_s > r_m$ :
\bi
\item $\gamma/\Gamma>2/3$  $\rightarrow$ $BS_{v<0}B$
\item $\gamma/\Gamma<2/3$ :
\bi
\item $r_s>|z_0|>r_m$  $\rightarrow$ $BS_hS_{v<0}B$
\item $r_s>r_m>|z_0|$  $\rightarrow$ $BS_hB$
\item $|z_0|>r_s>r_m$  $\rightarrow$ $BS_{v<0}B$
\ei
\ei
\ei
where $B$ denotes a bang pulse, $S_h$, $S_{v>0}$ and $S_{v<0}$ pulses along the magic plane, the line $(y=0,z>0)$ and the set $(y=0,z_0<z<0)$, respectively. %A bang is a very intense field of negligible duration inducing a perfect rotation of the spin without any relaxation effect.
\begin{figure}[htbp]
\centering\includegraphics[scale=0.5]{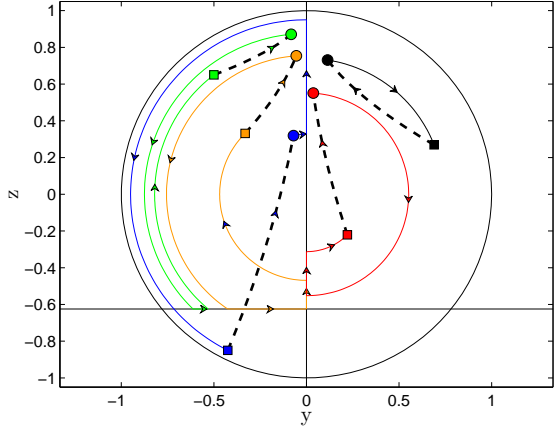}
\caption{\label{fig5} (Color online) Plot of the five possible optimal trajectories in the $(y,z)$- plane for the relaxation parameters $\Gamma=1.8$ and $\gamma=1$. The Ernst solution, which is a trajectory such that $M$ belongs to the Ernst ellipsoid (see the text for a definition), is represented in black. The circles and the squares indicate the position of the $S$ and $M$ points, respectively. The dashed lines display the detection period. A dynamical point of view of the trajectories is given by the different movies of the supplementary material.}
\end{figure}

\begin{figure*}[htbp]
\centering\includegraphics[scale=0.4]{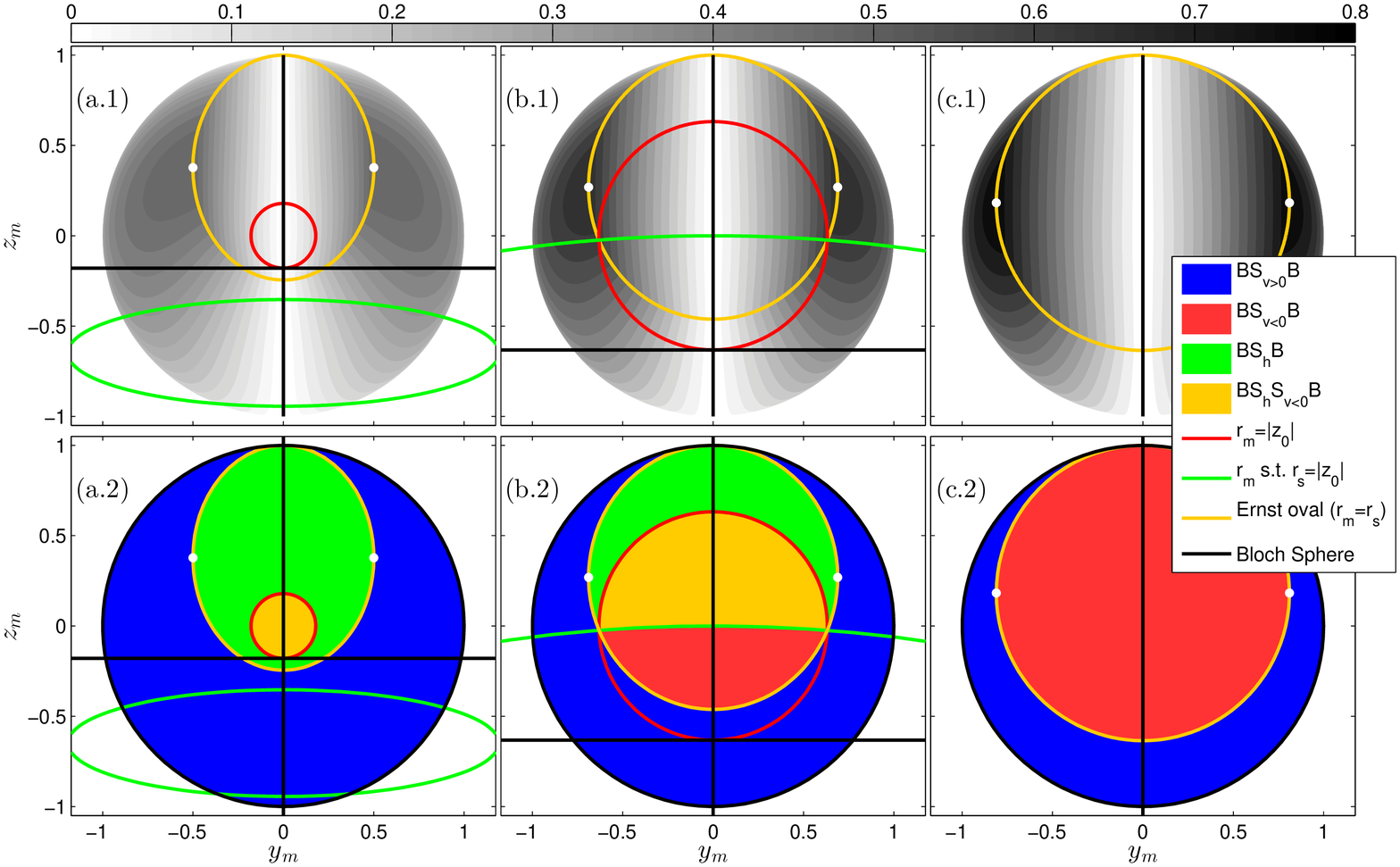}
\caption{\label{fig1} (Color online) (Top) Figure of merit surface $Q(y_m,z_m)$ associated with the three possible steady state syntheses. (Bottom) Steady state synthesis classifying the different types of control associated with each $M$ point of the Bloch ball. The parameters $(\Gamma,\gamma)$ are respectively taken to be $(1.90,0.5)$, $(1.80,1)$ and $(1.69,1.5)$, from left to right (See the text for details). Note that the same color code as in Fig. \ref{fig5} is used, except for the Ernst ellipsoid. The position of the $M$ point of the Ernst solution of coordinates $(y_m^{(E)},z_m^{(E)})$ (See Eq. (\ref{eqernst})) is represented by a white dot.}
\end{figure*}
From these different rules, we can identify the following sets of points such that $r_m=r_s$, $r_m=|z_0|$ and $r_s=|z_0|$ (the yellow, red and green curves in Fig. \ref{fig1}, respectively). For reasons which will become clear below, the set of points where $r_m=r_s$ is called in this work the \emph{Ernst ellipsoid} since it correspond to the set of $\delta$-pulses considered in the original paper by Ernst and Anderson \cite{ernstangle}. These different curves are the boundaries between different domains as can be seen in Fig. \ref{fig1}, a specific control law being associated with each domain.
%The points outside the Ernst oval need to increase the radius during the controlled period ($r_s<r_m$), while the radius will shrink for points inside. The points for which $r_s<r_m$ will use a control sequence of the form $BS_{v>0}$ (blue domain in Fig. \ref{fig1}). At the intersection of the sets $r_m<r_s$, $r_s=|z_0|$ and $r_m=|z_0|$ (red domain in Fig. \ref{fig1}), the control structure is of the form $BS_{v<0}B$. The points both inside the Ernst oval and the sphere defined by $r_m=|z_0|$ but outside $r_s=|z_0|$ use the $BS_hS_{v<0}B$ control sequence (light green domain in Fig. \ref{fig1}). The last region is the one inside the set defined by $r_s=r_m$ which does not intersect $r_m=|z_0|$ (green in Fig. \ref{fig1}). In this case, $r_s$ is larger than $z_0$, the magic plane is used to shrink the radius of the state vector.

The three boundaries can be explicitly computed in the space of $M$- points. For $r_m=|z_0|$, the curve is a circle of radius $|z_0|$ centered in $(0,0)$. The line of $M$ points such that $r_s=|z_0|$ corresponds to the backward free relaxation of the set $r_m=|z_0|$ during the detection period. For the Ernst ellipsoid, since the radius after the free evolution $r_s$ should be identical to $r_m$, we get in cartesian coordinates:
\begin{equation}\label{eqernstcart}
y_m^2\Exp{-\Gamma}+((z_m-1)\Exp{-\gamma}+1)^2-z_m^2-y_m^2=0.
\end{equation}
The transition from the case (a) to the case (b) of Fig. \ref{fig1} occurs when the set $r_s=z_0$ and the Ernst ellipsoid have only one point of intersection. This point belongs to the $z$- axis, with $z_s=z_0$. This is possible when the relaxation parameters satisfy $\Gamma=\frac{\gamma}{2}\frac{1-3\Exp{\gamma}}{1-\Exp{\gamma}}$. We observe a transition from the case (b) to the case (c) if the magic plane and the Bloch ball intersect in one point. This corresponds to $\Gamma=3\gamma/2$ (For $\Gamma\leq3\gamma/2$ the magic plane does not intersect the Bloch ball). Note that we have $\frac{\gamma}{2}\frac{1-3\Exp{\gamma}}{1-\Exp{\gamma}}\geq\frac{3\gamma}{2}$, which means that the two transition lines never cross in the $(\gamma,\Gamma)$- space and that only three different steady state syntheses exist (see Fig. \ref{fig4}). All the possible steady state syntheses have been identified, and the quality factor $Q$ can be evaluated now easily.
\section{Computation of the optimal steady state}\label{sec4}
In the unbounded case, all the rotations along the Bloch sphere are done instantaneously. Only the times spent on the magic plane and on the vertical line $y=0$ are different from zero. Such times can be computed straightforwardly by integrating the differential equations of the system \cite{lapertglaser,magic}. Plugging these different times in the quality factor $Q$, we can write the function $Q$ analytically (See the appendix \ref{appb} for the explicit derivation of the figure of merit $Q$). Figure \ref{fig1} shows the value of the figure of merit for three different cases. We observe that the maximum of $Q$ is reached for points associated with the control law $BS_{v>0}B$ (the blue domain in the steady state synthesis  of Fig. \ref{fig1}). The exact position of the maximum can be obtained by computing analytically the gradient $\nabla_{(y,z)} Q_{BS_{v>0}B}$. We find that the maximum of $Q_{BS_{v>0}B}$ belongs to the border of the set $BS_{v>0}B$, i.e. to the Ernst ellipsoid.
%\vspace{1.2cm}
%\begin{minipage}[c]{.475\linewidth}
\begin{figure}[htbp]
\centering\includegraphics[scale=0.55]{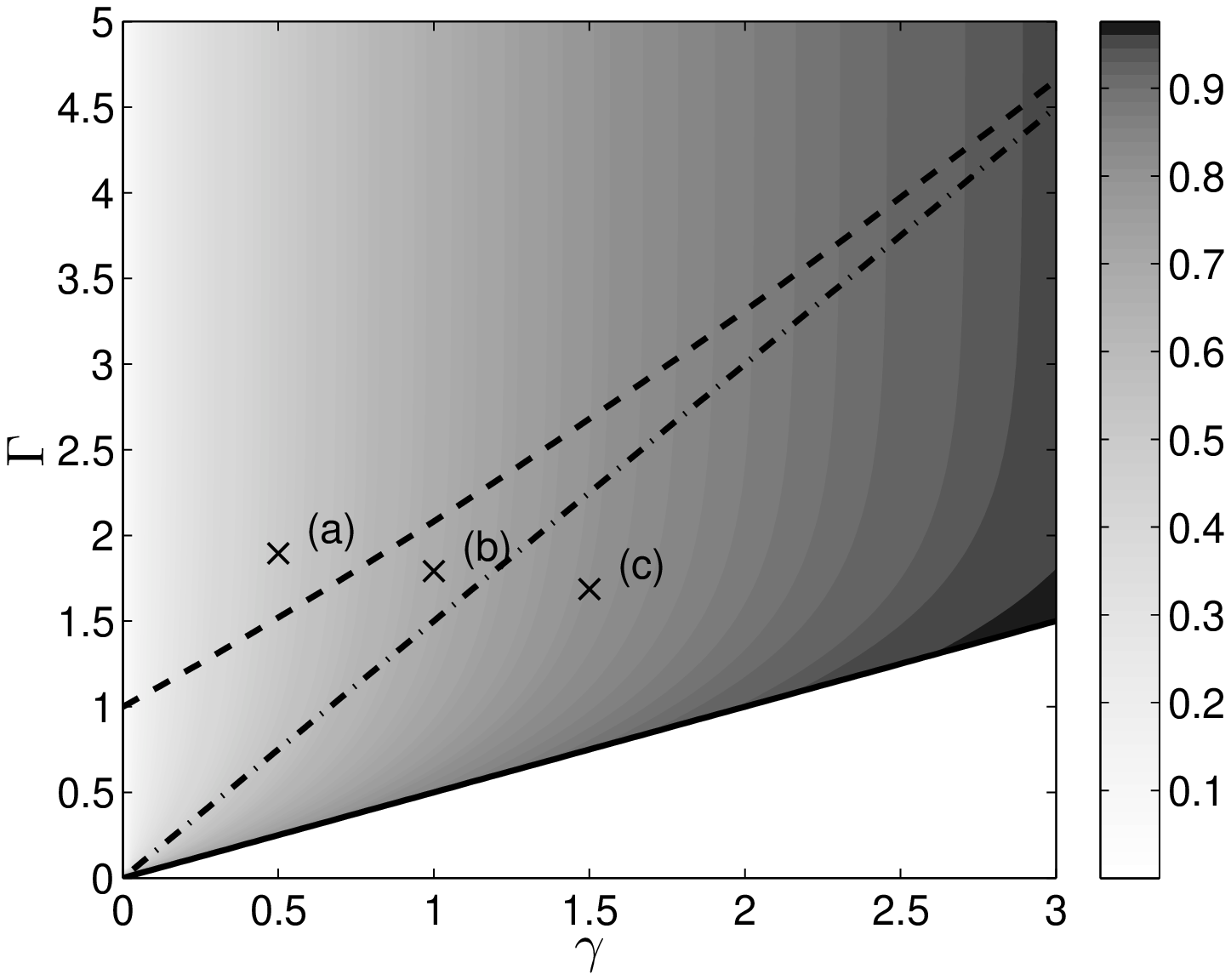}
\caption{\label{fig4} Maximum value of $Q$ given by Eq. (\ref{eqernst}) as a function of $(\gamma,\Gamma)$ (Ernst solution). The crosses denoted (a), (b) and (c) refer to the three cases of Fig. \ref{fig1}. The zone below the solid line is not physically relevant. The dotted dashed and dashed lines are the boundaries between the domains in which the steady state synthesis is of the form (a), (b) or (c).}
\end{figure}
Since the optimal solution belongs to Ernst ellipsoid, we could imagine that the Ernst angle solution should be optimal. Let us reformulate this known solution in our notation. The figure of merit on the Ernst ellipsoid is given by $Q=y_m$ since it requires only a bang pulse with $T_c=0$. Using Eq. (\ref{eqernstcart}), we can express $y_m$ as a function of $z_m$ and the derivative $d y_m(z_m)/d z_m$ is zero for $z_m^{(E)}=\frac{1}{1+\Exp{\gamma}}$ with a maximum given by:
\begin{equation}\label{eqernst}
y_m^{(E)}=\frac{\Exp{\Gamma}}{1+\Exp{\gamma}}\sqrt{\frac{\Exp{2\gamma}-1}{\Exp{2\Gamma}-1}},
\end{equation}
where the label $(E)$ refers to the Ernst solution. Figure \ref{fig4} displays the corresponding figure of merit $Q$ in terms of the relaxation parameters $\Gamma$ and $\gamma$. If we compute the $\theta$ angle of the bang pulse associated with this point and defined by $\theta=\theta_m^{(E)}-\theta_s^{(E)}$, we obtain:
\be
\cos\theta=\frac{\Exp{-\gamma}+\Exp{-\Gamma}}{1+\Exp{-\Gamma-\gamma}}.
\ee
This angle is nothing else that the well-known Ernst angle. This formula corresponds exactly to the formula (2.12) with a zero detuning of the original paper by Ernst and Anderson \cite{ernstangle}.
\section{Conclusion}\label{sec5}
In this work, we have shown that for unbounded pulse sequences, the Ernst angle solution is the optimal control law leading to the best SNR per unit time performance for an ensemble of homogeneous spin 1/2 particles. The Ernst angle solution was established in the sixties by considering only $\delta$- pulses. This result extends the optimality character of the Ernst angle to a more general framework, including singular arcs \cite{lapertglaser}, i.e. pulses with finite or zero control amplitudes. We have also introduced some geometric structures describing the periodic steady state process. The decisive advantage of our systematic approach over qualitative techniques is that the different constraints can be relaxed in order to describe more realistic experimental situations. In particular, this method can be generalized to the case of bounded control amplitudes, the use of crusher gradient pulses, to an inhomogeneous ensemble of spins with magnetic field broadening or to any other experimental constraint. On the basis of the same geometric approach, the steady state synthesis could be designed numerically in these different cases. In summary, this paper paves the way in a near future to a systematic analysis of the optimization of the SNR per unit time of quantum systems. Work is in progress on these open questions.\\ \\
\noindent\textbf{ACKNOWLEDGMENT}\\
S.J. Glaser acknowledges support from the DFG (GI 203/7-1), SFB 631 and the BMBF FKZ 01EZ114 project. M. Lapert acknowledges support from Alexander von Humboldt Stiftung. D. Sugny and S. J. Glaser acknowledge support from the PICS program of the CNRS and from the QUAINT coordination action (EC FET-Open). E. Ass\'emat is supported by the Koshland Center for basic Research.

\appendix

\section{Time-optimal control of the Bloch equation}\label{appa}
We explicitly show in this appendix how to construct  the different sheets of the figure of merit surface $Q$. For sake of completeness, we briefly outline some results of Ref. \cite{magic} which are used throughout the paper.

The dynamics of the system is governed by the following Bloch equation written in the $(y,z)$- plane:
\be
\lgr\ba{rl}
\dot{y}&=-\Gamma y -u z\\
\dot{z}&=\gamma(1-z)+u y
\ea\rd.
\ee
%The magic plane can be defined as the set of points for which the shrinking of the modulus of the magnetization vector is the fastest.
The time-optimal control law can be completely described by two geometric objects of the Bloch ball, the magic plane and the steady-state ellipsoid \cite{magic}. In the unbounded case, there is a complete control on the two angular coordinates. In other words, this means that all the rotations along the Bloch sphere can be made in an arbitrarily short time. The geometric nature of the optimal dynamics can thus be revealed by the evolution of the radial coordinate $r=\sqrt{y^2+z^2}$.

In polar coordinates, $y=r\cos\theta,~z=r\sin\theta$, the dynamics is ruled by the differential equations:
\be
\lgr\ba{rl}
\dot{r}&=-\Gamma r\cos^2\theta+\gamma\sin\theta-\gamma r\sin^2\theta, \\
\dot{\theta}&=\gamma\cos\theta(\frac{1}{r}-\sin\theta)+\Gamma\cos\theta\sin\theta+u.\nonumber
\ea\rd
\ee
Following Ref. \cite{magic}, the geometric objects are determined by computing the zeros of the angular derivative of the radial speed:
\begin{equation}\label{eqder}
\frac{d\dot{r}}{d\theta}=\frac{\sqrt{y^2}}{\sqrt{y^2+z^2}}\big(2\Gamma z+\gamma-2\gamma z\big).
\end{equation}
In particular, the points of the steady state ellipsoid are the zeros of this derivative. The corresponding ellipse divides therefore the $(y,z)$- plane into two domains, the inner and the outer ones being respectively the set of points where the radius grows and shrinks. Other characteristic points of the derivatives are given by its maxima and minima. Using Eq. (\ref{eqder}) and the second derivative of $r$ with respect to $\theta$ \cite{magic}, it can be shown that the shrinkage of the radius $r$ is maximum on the magic plane of equation $z=z_0=-\frac{\gamma}{2(\Gamma-\gamma)}$. Note that for some values of $\Gamma$ and $\gamma$, the magic plane does not intersect the Bloch ball. The $z$- axis is also obtained as solution. This line can be divided into three parts. The part $z>0$ corresponds to the set of maximum growth of the $r$- component. The part $0>z>z_0$ is the set of maximum shrinking and the set $z<z_0$ is never optimal, neither for growing nor for shrinking \cite{magic}. The magic plane and the $z$- axis are used in the steady-state synthesis.
\section{Explicit derivation of the figure of merit surface $Q$}\label{appb}
The explicit derivation of the figure of merit surface $Q$ requires the computation of the time of travel along the magic plane and the $z$- axis. It is straightforward to show that, for the vertical line of equation $y=0$, the associated control field is of the form $u_{y=0}(t)=0$, while for the magic plane, starting from $\dot{z}=0$, we arrive at $u_{z=z_0}(t)=-\gamma (1-z_0)/y(t)$. More precisely, we need to compute the time to move from $y_1$ to $y_2$ along the magic plane and the time to move from $z_1$ to $z_2$ along the line $y=0$. In the two cases, replacing $u$ and $y$ or $z$ by their respective expressions lead to:
\begin{equation}
\dot{y}=-\Gamma y + \gamma(1-z_0)z_0/y
\end{equation}
for the magic plane and
\begin{equation}
\dot{z}=\gamma(1-z)
\end{equation}
for the $z$- axis. Both differential equations can be easily solved and we find the following times of travel:
\be
\lgr\ba{rl}\label{eq4}
T_{z=z_0}(y_1\rightarrow y_2)&=\frac{1}{2\Gamma}\ln \frac{ 4\left( \Gamma-\gamma \right) ^2y_1^2+\gamma^2 \left( 2\Gamma-\gamma \right) }{4\left( \Gamma-\gamma \right) ^2y_2^2+\gamma^2 \left( 2\Gamma-\gamma \right)}\\
T_{y=0}(z_1\rightarrow z_2)&=\frac{1}{\gamma}\ln\frac{1-z_1}{1-z_2}.
\ea\rd
\ee
From Eq. (\ref{eq4}), it is straightforward to write down the expression of the figure of merit surface $Q$ for the four different kinds of control structure introduced in the main text. For the trajectories using only the set $y=0$, $BS_{v<0,v>0}B$ (see Fig. 3 of the main text), it is sufficient to know the radius $r_s$ associated with $(y_s,z_s)$ and the radius $r_b=\sqrt{y_s^2\Exp{-2\Gamma}+((z_s-1)\Exp{-\gamma}+1)^2}$ obtained after free relaxation. The figure of merit $Q$ can be expressed as:
\begin{equation}\label{eq5}
\ba{rl}
Q_{BS_v<0B}&=\frac{y_s}{\sqrt{1+\frac{1}{\gamma}\ln\frac{1+\sqrt{(y_s\Exp{-\Gamma})^2+((z_s-1)\Exp{-\gamma}+1)^2}}{1+\sqrt{y_s^2+z_s^2}}}}\\
Q_{BS_v>0B}&=\frac{y_s}{\sqrt{1+\frac{1}{\gamma}\ln\frac{1-\sqrt{(y_s\Exp{-\Gamma})^2+((z_s-1)\Exp{-\gamma}+1)^2}}{1-\sqrt{y_s^2+z_s^2}}}}.\\
\ea
\end{equation}
The trajectory $BS_hB$ using only the magic plane  (see Fig. \ref{fig1}) consists in moving from the radius $r_b$ to $r_s$ which means moving from $y_1=\sqrt{r_b^2-z_0^2}$ to $y_2=\sqrt{r_s^2-z_0^2}$. We get:
\begin{equation}\label{eq6}
Q_{BS_hB}=\frac{y_s}{\sqrt{1+\frac{1}{2\Gamma} \ln \frac {4(y_s^2\Exp{-2\Gamma} +\left(  \left( z_s-1 \right) \Exp{-\gamma}+1 \right) ^2)\Gamma \left( \Gamma-\gamma \right) +\gamma^2}{4 \left(y_s^2+z_s^2 \right) \Gamma \left( \Gamma-\gamma \right) +\gamma^2}}}
\end{equation}
The last possible trajectory is of the form $BS_hS_{v<0}B$ (see Fig. \ref{fig1}). It consists of a rotation from the point $(y_b,z_b)$ to the magic plane, followed by a travel along this plane until the vertical axis $y=0$, and a last travel along this vertical line from $z_0$ to $r_s$. Again it is straightforward to show that:
\begin{equation}\label{eq7}
Q_{BS_hS_{v<0}B}=\frac{y_s}{\sqrt{1+\frac{1}{2\Gamma} \ln \frac {4(y_s^2\Exp{-2\Gamma} +\left(  \left( z_s-1 \right) \Exp{-\gamma}+1 \right) ^2)\Gamma \left( \Gamma-\gamma \right) +\gamma^2}{\gamma^2}+\frac{1}{\gamma}\ln\frac{\frac{2\Gamma-\gamma}{2(\Gamma-\gamma)}}{1+\sqrt{y_s^2+z_s^2}}}}.
\end{equation}
Equations (\ref{eq5}), (\ref{eq6}) and (\ref{eq7}) give the four sheets of the $Q$ surface. Note that $Q$ is a continuous but not smooth surface. A contour plot of this surface is represented in Fig. \ref{fig1} for different values of the relaxation parameters $\Gamma$ and $\gamma$.

\end{document}